\documentclass{article}

\usepackage{microtype}
\usepackage{graphicx}
\usepackage{subcaption}
\usepackage{booktabs} 
\usepackage{enumitem}

\PassOptionsToPackage{hyphens}{url}\usepackage{hyperref}
\usepackage{hyperref}

\usepackage[accepted]{icml2024}

\usepackage{amsmath}
\usepackage{amssymb}
\usepackage{mathtools}
\usepackage{amsthm}
\usepackage[subtle, tracking=normal]{savetrees}
\usepackage{hyperref}
\usepackage{url}
\usepackage{xspace}
\usepackage{enumitem}
\usepackage{graphicx}

\usepackage[capitalize,noabbrev]{cleveref}

\theoremstyle{plain}

\theoremstyle{definition}

\theoremstyle{remark}

\usepackage[textsize=tiny]{todonotes}

\newcommand*{\sn}{\textsc{ZkAudit}\xspace}
\newcommand*{\snt}{\textsc{ZkAudit-T}\xspace}
\newcommand*{\sni}{\textsc{ZkAudit-I}\xspace}

\newcommand{\minihead}[1]{{\vspace{.5em}\noindent\textbf{#1.} }}

\icmltitlerunning{Trustless Audits without Revealing Data or Models}

\begin{document}

\twocolumn[
\icmltitle{Trustless Audits without Revealing Data or Models}



\icmlsetsymbol{equal}{*}

\begin{icmlauthorlist}
\icmlauthor{Suppakit Waiwitlikhit}{stanford}
\icmlauthor{Ion Stoica}{berkeley}
\icmlauthor{Yi Sun}{chicago}
\icmlauthor{Tatsunori Hashimoto}{stanford}
\icmlauthor{Daniel Kang}{uiuc}
\end{icmlauthorlist}

\icmlaffiliation{stanford}{Stanford University}
\icmlaffiliation{berkeley}{UC Berkeley}
\icmlaffiliation{chicago}{University of Chicago}
\icmlaffiliation{uiuc}{UIUC}

\icmlcorrespondingauthor{Daniel Kang}{ddkang@illinois.edu}

\icmlkeywords{Machine Learning, ICML}

\vskip 0.3in
]



\printAffiliationsAndNotice{\icmlEqualContribution} 

\begin{abstract}

There is an increasing conflict between business incentives to hide models and
data as trade secrets, and the societal need for algorithmic transparency. For
example, a rightsholder wishing to know whether their copyrighted works have
been used during training must convince the model provider to allow a third
party to audit the model and data. Finding a mutually agreeable third party is
difficult, and the associated costs often make this approach impractical.

In this work, we show that it is possible to simultaneously allow model
providers to keep their model weights (but not architecture) and data secret
while allowing other parties to trustlessly audit model and data properties. We
do this by designing a protocol called \sn in which model providers publish
cryptographic commitments of datasets and model weights, alongside a
zero-knowledge proof (ZKP) certifying that published commitments are derived
from training the model. Model providers can then respond to audit requests by
privately computing any function $F$ of the dataset (or model) and releasing the
output of $F$ alongside another ZKP certifying the correct execution of $F$. To
enable \sn, we develop new methods of computing ZKPs for SGD on modern neural
nets for simple recommender systems and image classification models capable of
high accuracies on ImageNet.  Empirically, we show it is possible to provide
trustless audits of DNNs, including copyright, censorship, and counterfactual
audits with little to no loss in accuracy.

\end{abstract}

\section{Introduction}
\label{sec:intro}

As ML models become more capable, businesses are incentivized to keep
the model weights and datasets proprietary. For example, Twitter recently
released their algorithm but not the model weights \cite{twitter2023alg}, and
many LLM providers only provide access via APIs. On the other hand, there is
also an increasing societal need for transparency in the model and data behind
these APIs: closed models harm transparency and trust \cite{karissa2023what}.

As ML has become more capable and is being deployed in sensitive settings,
models and data are increasingly kept proprietary. This is because data and
model weights are increasingly treated as trade secrets. Furthermore, the data
cannot be released in many circumstances, such as in medical settings.
Unfortunately, keeping models and training data secret harms transparency and
trust. Users may have specific requirements when using gated models, such as
avoiding copyrighted material. Model providers may lose customers due to this
lack of transparency.

To address this, we want to perform audits where model providers and users agree
on specific properties to test for the ML training procedure and models. For
example, the property that the training dataset contains no copyrighted content
or a recommender system is not censoring items (e.g., tweets). An audit would
ideally release results for exactly these properties and nothing else.

Currently, there are three methods of performing such audits on modern ML
methods. One method is to release the data, random seed, and final model
weights: the user can replay training. However, this procedure does not keep the
data and weights hidden. Another method is multi-party computation (MPC), which
allows several parties to participate in a computation while keeping data
hidden. Unfortunately, MPC requires all participants to participate honestly,
but an audit presupposes a lack of trust between the model provider and users.
MPC that handles malicious adversaries is extremely bandwidth intensive: a
back-of-the-order calculation suggests the training procedure for an
\emph{8-layer CNN} may take up to \emph{5 petabytes} of communication, which
would cost \$450,000 in cloud egress fees \cite{pentyala2021privacy}. Finally, a
trusted third party (TTP) could perform the audit, but TTPs are rarely
practical. The TTP must have access to trade secrets, which model providers wish
to keep secret. Furthermore, audits are expensive, requiring highly specialized
expertise (deep understanding of ML training), and strong security (to avoid
leaking trade secrets). In many cases, no viable TTPs are trusted by both model
providers and users.

Prior work has proposed using zero-knowledge proofs to perform audits to address
this issue \cite{kroll2015accountable, shamsabadi2022confidential}.
Zero-knowledge proofs allow a prover to prove properties about their data (e.g.,
training data or model weights) without revealing the data itself. However, none
of this prior research extends to modern ML methods in the form of deep neural
networks (DNNs).

In this work, we develop an auditing procedure \sn that can perform audits
\emph{without third parties and without any assumptions of trust} (i.e.,
trustlessly) on modern DNNs. \sn, via zero-knowledge proofs, allows a model
provider to selectively reveal properties of the training data and model without
a TTP such that any party can verify the proof after the fact (i.e., the audit
is non-interactive). Importantly, these guarantees are \emph{unconditional},
providing security against malicious adversaries with only standard
cryptographic assumptions.

\sn consists of two steps: \snt and \sni. In \snt, the model provider trains a
model and publishes a commitment (e.g., hash) of their dataset, model weights,
and a zero-knowledge proof that proves the weights were generated by training on
the committed dataset. Then, in \sni, the user provides an arbitrary audit
function $F$. The model provider executes $F$ on the \emph{same} weights and
dataset used in training and provides a zero-knowledge proof of the
execution. The zero-knowledge proof guarantees that $F(\textrm{data},
\textrm{weights})$ was executed on the hidden model/weights and evaluated
honestly. For example, $F$ could check whether a training set contains
copyrighted data or whether a social media provider is shadowbanning posts. The
model provider can trustlessly evaluate $F$ by using prior work to generate
zero-knowledge proofs for inference \cite{lee2020vcnn, weng2022pvcnn,
feng2021zen, kang2022scaling}.


To enable \sn, we leverage recent development in cryptographic techniques known
as ZK-SNARKs (zero-knowledge succinct non-interactive argument of knowledge).
ZK-SNARKs allow a prover to produce a proof that an arbitrary computation
happened correctly (Section~\ref{sec:background}). However, ZK-SNARKs are
incredibly costly: it can take up to days to prove the execution of the forward
pass on even toy models \cite{lee2020vcnn, weng2022pvcnn, feng2021zen}. Only
recently has it become possible to produce proofs of the forward pass on
real-world models \cite{kang2022scaling}. However, no existing work can compute
the backward pass necessary for gradient descent, a requirement for proofs of
training.

To produce proofs of training, we extend recent work to compute the backward
pass of real-world DNNs. Our work enables model providers to produce proofs of
full stochastic gradient descent on private data. Doing so requires overcoming
several challenges: prior work uses integer division and \texttt{int8} precision
for efficient forward pass computation. Unfortunately, training with these
settings is not amenable to achieving high accuracy. We provide methods of
embedding stochastic gradient descent with rounded division and variable
fixed-point precision, and show that training in fixed-point can achieve high
accuracy.



On commodity hardware, \sn can produce audits of image classification systems
and simple recommender systems with little to no loss in accuracy on a range of
real-world datasets (medical datasets and standard benchmark datasets). The cost
of auditing a recommender system and image classification system can be as low
as \$10 and \$108, respectively, showing the practicality of our work. Achieving
these low costs requires all of our optimizations: training would suffer
dramatic losses in accuracy or not proceed without them.



\section{Background on ZK-SNARKs}
\label{sec:background}

%
%
%
%

\minihead{ZK-SNARKs}
ZK-SNARKs are a cryptographic primitive that allows a \emph{prover} to produce a
proof $\pi$ that some function $F(x, w)$ was computed correctly, where $x$ is
public and $w$ is private. Given $\pi$, a verifier can check that the prover
computed $F$ correctly without access to $w$.

ZK-SNARKs have several amazing properties. First, they are \emph{succinct},
i.e., small in the size of the input. Second, they are \emph{non-interactive},
meaning the prover and verifier need not interact beyond $\pi$. Third, they are
\emph{knowledge sound}, which means that a computationally bounded prover cannot
generate proofs for incorrect executions.  Fourth, they are \emph{complete},
meaning that proofs of correct execution verify (often unconditionally).
Finally, they are \emph{zero-knowledge}, which means $\pi$ reveals nothing about
the private inputs beyond what the output and public inputs contain.

Although ZK-SNARKs allow arbitrary computation to be proved, ZK-SNARKs require
computation to be expressed in specific ways. The cryptography community has
provided several such ways of expressing computations, including R1CS
\cite{groth2016size} and Plonk \cite{gabizon2019plonk}.

Unfortunately, naively expressing computations can result in highly inefficient
proof generation. The specification of the computation and proving systems can
jointly result in \emph{three orders of magnitude} or more differences in
proving times.


\minihead{Representing computation}
We describe salient details of representing computation in ZK-SNARKs. Although
other works describe relevant details, it is critical to understand the basic
building blocks and costs associated with computation in ZK-SNARKs to understand
our optimizations.

In this work, we leverage arithmetic intermediate representations (AIRs),
represented by a 2D grid $x_{ij}$ of values. We denote the number of rows as $R$
and columns as $C$. Due to the construction of ZK-SNARKs, the $x_{ij} \in
\mathbb{F}_q$ for some large prime $q$. In particular, arithmetic is done
\emph{in the finite field}, so standard operations such as division are not
natively possible.

Logically, there are three ways to constrain values on the grid:
\begin{enumerate}
  \item Constraining two values to be equal: $x_{ij} = x_{i' j'}$.

  \item Constraining a subset of a row to be in a pre-defined table: $(x_{i
  j_1}, ..., x_{i j_k}) \in \{ (t_1, ..., t_k) \} = T_m$ for some table $T_m$.
  $T_m$ is called a lookup table.

  \item A polynomial constraint on a grid row: $f_l(x_{i1}, ..., x_{iR}) =
  0$ for some polynomial $f_l$.

\end{enumerate}
We provide an example of using polynomial constraints to implement integer
division in Section~\ref{sec:backprop}, and \citet{kang2022scaling} provide
other examples.

In general, the costs increase with the number of rows ($R$), number of columns
($C$), maximum degree of the polynomial constraints $f_l$, and number of lookup
tables ($T_m$). Furthermore, the number of rows must be a power of two.
Given a representation of a computation in an AIR, we can produce a concrete
ZK-SNARK proof by using a proving system such as halo2 \cite{halo2}.

We provide an extended discussion of ZK-SNARKs in Appendix~\ref{sec:apx-snarks},
including examples of using AIRs and how to compile ZK-SNARKs.

\section{\sn: Private Audits of ML}
\label{sec:protocol}


\minihead{Protocol}
We describe \sn when given access to verified randomness, a public source of
timestamped, verified random bits. The need for verified random bits can be
removed with the slightly stronger assumption of a random oracle hash
function, which we describe in Appendix~\ref{apx:no-verified-random}.
Throughout, we assume access to a binding and hiding commitment scheme, in which
the trainer commits to the training data and cannot change the values later. The
commitment scheme can be implemented in practice by publicly releasing hashes of
the data and weights.

The first part of \sn (\snt) proves that the trainer honestly trained a model
with known architecture but hidden weights on a hidden dataset. To do so, the
trainer commits to the data, commits to a training order, and produces ZK-SNARKs
for SGD from a randomly initialized or public pre-trained model:
\begin{enumerate}
  \item The trainer commits to a dataset $\{ (x_1, y_1), ..., (x_n, y_n) \}$,
  producing commitments $[c_1, ..., c_n ]$. The commitments are
  ordered lexicographically, and the trainer publicly posts the commitments.

  \item The trainer uses a verified source of randomness to generate a traversal
  ordering of the dataset (see Appendix~\ref{apx:protocol} for why this is
  desired in some circumstances).

  \item The trainer computes ZK-SNARKs of the SGD process, one batch at a time,
  using the traversal ordering. To do so, it computes the ZK-SNARK of the
  forward pass of any frozen layers, the forward and backward pass of any
  layers being updated, and the weight update procedure. This can be done in one
  or more ZK-SNARKs.

  \item The trainer publishes the ZK-SNARKs of SGD and the commitment to the
  model weights at the end of training.
\end{enumerate}

The second part of the protocol (\sni) computes the zero-knowledge proof(s) for
the audit function itself. Given the commitments to the dataset and final
weights, the user sends an audit function $F(\textrm{data}, \textrm{weights})$.
The trainer then computes a ZK-SNARK of the audit function and publishes it
(along with the output of $F$). For example, the audit may be that a recommender
system is not censoring social media content. The model trainer must also hash
the weights in the zero-knowledge proof to ensure trained weights from \snt are
consistent with the weights in \sni.


\minihead{Security analysis}
\sn has the following (informal) properties: 1) the trainer cannot ``cheat'' in
training or computing the audit function, and 2) the verifier learns nothing about
the training data and model weights aside from the output of the audit function.
We can formalize these properties as knowledge soundness and zero-knowledge.

We provide a formal analysis of security (knowledge soundness and
zero-knowledge) in Appendix \ref{apx:protocol} and provide an informal
analysis here. For our security analysis, we assume two standard cryptographic
primitives: a cryptographically secure hash function (informally, one that is
secure against collisions) \cite{menezes2018handbook} and ZK-SNARKs
\cite{bitansky2017hunting}.  It is standard to denote the security of these
primitives with a parameter $\lambda$.  Informally, the security parameter
controls the probability that an adversary can ``break'' the protocol.

Denote the dataset size as $D$ and the number of stochastic gradient steps as
$T$. Then, the prover produces at most $D + 4 T$ hashes, commitments, and
ZK-SNARKs. The security of each hash and ZK-SNARK follows directly from the
primitives. 

By the union bound, in order to achieve a security parameter of $\lambda$ for
\sn, we must choose the hash function and ZK-SNARK parameters so they have at least
$(D + 4 T) \lambda$ bits of security.

\minihead{Security of ZK-SNARKs}
\sn's security rests on the security of the underlying ZK-SNARK proving backend
that is used (halo2~\cite{halo2} in this work). Our ZK-SNARKs can be constructed
via KZG commitments \cite{kate2010constant} or inner-product arguments (IPA)
\cite{bunz2021proofs}.  In the KZG version, we require a
\emph{structured-reference string} (SRS) that is universal to \emph{all} audit
functions. Namely, the SRS need only be generated once in a secure manner. To do
so, we can use the already-generated SRS, which was generated using a perpetual
trusted setup in which many parties participate (over 75 at the time of writing)
\cite{powersoftau}. Only a single party needs to be honest for the setup to be
secure.  IPAs do not require any trusted setup.

\minihead{Limitations}
Although \sn provides computational security against malicious adversaries and
traversal ordering attacks, it has two major limitations. First, it does not
protect against data poisoning attacks, in which a malicious attacker can
manipulate the data itself \cite{steinhardt2017certified}. Second, while \sn
does not reveal the weights, it does reveal the model architecture. We view
addressing these limitations as exciting future research.

\section{Computing ZK-SNARKs for Gradient Descent}
\label{sec:backprop}

We now describe our method and optimizations for computing gradient descent
within a ZK-SNARK.
%
Unlike in the computation of the forward pass, the input to gradient descent
is both the input data and the model weights. The output is an updated set of
weights.  Formally, for an input $x$ and weights $w$, gradient descent computes
$w' = G(x, w)$, where $w'$ is the updated set of weights. One standard method of
performing gradient descent is to compute the forward pass, compute the backward
pass, and update the weights by scaling the gradients by the learning rate.

Prior work has optimized the forward pass for \texttt{int8} inference in
ZK-SNARKs \cite{kang2022scaling}. In this work, we extend this prior work by
showing how to compute the backward pass in a ZK-SNARK. We further optimize
gradient descent by designing a high-performance softmax in ZK-SNARKs and
operating in fixed-point arithmetic.

We first observe that the backward pass can often be expressed in structurally
similar computation as the forward pass. For example, the backward pass of a
convolution can also be expressed as a convolution (with different inputs).

However, several significant differences between inference and training
necessitate changes.

\minihead{Rounded division and fixed-point}
Training requires more precise arithmetic than inference. For efficiency, prior
work \citet{kang2022scaling} uses the floor function for \texttt{int8}
arithmetic, which would result in poor accuracy for training. To understand why,
consider the update formula for SGD:
$
w' = w + \eta \cdot \Delta w
$
Typically, the learning rate $\eta$ is small (e.g., 0.01). When using lower
precision, the multiplication by $\eta$ can be imprecise, leading to poor
accuracy.

Thus, in order to compute ZK-SNARKs for gradient descent, we introduce two
techniques: rounded division in finite field ZK-SNARK constraints and variable
precision fixed-point arithmetic. Both techniques increase the accuracy of
training.

We first implement rounded division in finite fields with polynomial constraints.
As we show (Section~\ref{sec:eval}), using rounded division can improve accuracy
by up to 11\% compared to standard integer division. We first describe how
to implement standard integer division (which rounds towards zero). Suppose that
$a, b, c, r$ are all positive. If
$
b = \left\lfloor \frac{a}{c} \right\rfloor
$
then we have that the following constraint holds
\begin{align}
a = b \cdot c + r
\label{eq:int-div-poly}
\end{align}
where $0 \leq r < c$. To implement standard integer division, we first
assume that $0 \leq b, c < 2^N$ for some $N$. We can then use the polynomial
constraint in Equation~\ref{eq:int-div-poly}, constrain that $b, c, r \in \{0,
..., 2^N - 1 \}$, and that $c - r \in \{ 0, ..., 2^N - 1 \}$. Constraining that
$c - r \in \{ 0, ..., 2^N - 1 \}$ is equivalent to the constraint that $c > r$.

To implement rounded division, consider $a, b, c, r$ all positive
integers. As before, we assume that $0 \leq b, c < 2^N$ for some $N$. Let
$
b = \left[ \frac{a}{c} \right].
$
Then, the following constraints specify rounded division
\begin{align}
2 a + c = 2 c \cdot b + r
\label{eq:round-div-poly}
\end{align}
where $0 \leq r < 2c$. This follows because
\[
b = \left\lfloor \frac{2a + c}{2c} \right\rfloor = \left\lfloor \frac{a}{c} +
\frac{1}{2} \right\rfloor.
\]
We can use similar constraints: Equation~\ref{eq:round-div-poly}, a constraint
that $b, c \in \{0, ..., 2^N - 1 \}$, and a constraint that $2c - r \in \{0,
..., 2^{2N} - 1 \}$. Although this requires a lookup table of size $2^{2N}$, we
can implement rounded division, which is critical for training.

We further implement variable precision fixed-point arithmetic to allow
trade-offs between accuracy and computation. Fixed-point arithmetic
approximates real numbers by $\hat{x} = \textrm{Round}(x \cdot \textrm{SF})$,
where SF is the scale factor.
Since we use lookup tables to allow for non-linearities and fixed-point
rescaling, more precision (i.e., a larger scale factor) results in larger lookup
tables. This directly results in higher proving times but allows the model
trainer to decide which precision level to use.

\minihead{Softmax}
In order to perform classification, we designed a high-performance
softmax in ZK-SNARKs. To understand the difficulties of implementing softmax
with finite field operations, recall the explicit formula:
$y_i = \frac{e^{x_i}}{\sum_j e^{x_j}}.$
Denote $s = \sum_j e^{x_j}$ and $\hat{e} = [e^{x_i}]$.
Naively computing the softmax using the operations present would compute the
exponential with a lookup table \emph{in the standard units by scaling by the
scale factor}, sum $e^{x_i}$, then divide by $s$. However, we must address three
challenges when using fixed-point arithmetic to compute the softmax: underflow,
precision, and range.

To understand these issues, consider a toy example where $x = [ \ln \frac{1}{2},
0 ]$, so $\hat{e} = [ \frac{1}{2}, 1 ]$, $s = \frac{3}{2}$, and $y = [
\frac{1}{3}, \frac{2}{3} ]$ in full precision. Consider using a scale factor of
1000 for simplicity. The first issue arises in dividing by $s$: in the scaled
units, $\hat{e} = [ 500, 1000]$, so $s = 1500$. However, \emph{when dividing in
the scaled units}, $y = [0, 1]$. Thus, naive computation would result in a
substantial loss in precision (underflow).

We can address the issue of underflow by dividing $s$ by the scale factor.
However, this results in $s = 2$ and $y = [250, 500]$ in scaled units or
$[\frac{1}{4}, \frac{1}{2}]$. This results in a relative error of \emph{33\%} in
$y$, a substantial degradation in accuracy. In order to address this, we
can scale $e^{x_i}$ by the scale factor again and not divide $s$ by the scale
factor.


Finally, we use a standard trick to increase the numeric stability of the
softmax. Since the softmax is shift-invariant, subtracting the maximum value
results in a smaller range of outputs of the exponentiation. 

To compute the maximum of a vector, we can compute the pairwise maximum
sequentially. In order to compute the pairwise maximum $c = \max(a, b)$
efficiently, we can use the following constraints. First, we constrain that $c$
is one of $a$ or $b$ by using the polynomial constraint
$
(c - a) \cdot (c - b) = 0.
$
We then constrain that $c - a, c - b \in [0, ..., 2^N)$, where $2^N$ is the size
of our lookup table. This enforces that $c \geq a, b$.

\section{Evaluation of \snt}
\label{sec:eval}

We now evaluate \snt, including the performance of performing SGD in ZK-SNARKs,
the end-to-end accuracy and costs of \snt, and the effect of our optimizations.
Because verification of ZK-SNARKs is cheap (10ms per proof), we focus on the
cost of proving, which far dominates, and the accuracy (since there are
potential degradations when discretizing).

We benchmarked SGD and \snt on image classification and a recommender system on
Movie Lens \cite{harper2015movielens}. For the image classification tasks, we
used a variety of MobileNet v2 configurations. The MobileNet configurations are
denoted by the depth multiplier and input resolution, so MobileNet (1.0, 224) is
a MobileNet v2 with a depth multiplier of 1.0 and an input resolution of
224$\times$224. For the recommender system, we used a small model based on the
DLRM model from Facebook \cite{naumov2019deep}. The complete configuration is in
Appendix~\ref{apx:eval-setup}.

To generate the cost estimates, we multiplied the total computation time by the
cost of using a cloud computing platform (AWS) to perform the computation
(Appendix~\ref{apx:eval-setup}). We further conducted experiments on CIFAR-10 in
the Appendix.

\subsection{Performance of SGD}
We first investigated the performance of embedding the computation of a single
SGD step in a ZK-SNARK. We measure the proving time, verification time, and
proof size.

\begin{table}[t!]
\centering
  \centering
  \begin{tabular}{lll}
    Scale factor & Proving time & Verification time \\
    \hline
    $2^{12}$ & 47.5 s  & 10.0 ms \\
    $2^{13}$ & 87.8 s  & 9.9 ms \\
    $2^{14}$ & 167.0 s & 9.9 ms \\
    $2^{15}$ & 328.3 s & 9.8 ms \\
  \end{tabular}
  \caption{
  Proving and verification time of SGD on scale factors for image classification
  on a single image (MobileNet v2 (1.0, 224)). The proof size was 9.03 kb for
  all configurations.}
  \label{table:sgd-timing}
\end{table}

\begin{table}[t!]
  \centering
  \begin{tabular}{lll}
    Scale factor & Proving time & Verification time \\
    \hline
    $2^{11}$ & 3.16 s  & 6.2 ms \\
    $2^{12}$ & 5.54 s  & 6.1 ms \\
    $2^{13}$ & 10.49 s & 6.3 ms \\
    $2^{14}$ & 23.79 s & 6.0 ms \\
  \end{tabular}
  \caption{Proving time and verification time of SGD on a variety of scale
  factors for a recommender system (single example). The proof size was 4.6 kb for all
  configurations.}
  \label{table:sgd-time-recsys}
\end{table}

\begin{table}[t!]
  \centering
  \begin{tabular}{llll}
    Method & Proving lower bound \\
    \hline
    Zen    & 200,000* s \\
    vCNN   & 172,800 s  \\
    pvCNN  & 31,011* s  
  \end{tabular}
  \caption{Estimated lower bounds for proving times of prior work for image
  classification on a \emph{single} image. We exclude zkCNN since the authors
  explicit state that they are unable to compute the softmax
  \cite{liu2021zkcnn}, so are unable to compute proofs of SGD. 
  }
  \label{table:lower-bounds}
\end{table}

We show results for image classification in Table~\ref{table:sgd-timing} and for
the recommender system in Table~\ref{table:sgd-time-recsys}. The verification
costs for ZK-SNARKs of SGD are incredibly low: as low as 6.0 ms.  The proving
times range from 26s to 328s for image classification and 2 s to 48s for the
recommender system. Furthermore, none of the prior work implements the softmax
operation, making SGD infeasible with this work. Nonetheless, comparing the
proving times of our work to the proving times of just the arithmetic
operations of prior work shows that our work is at least 95$\times$ faster
(Table~\ref{table:lower-bounds}).


\subsection{End-to-End Accuracy and Costs}
We then benchmarked the end-to-end accuracy of fine-tuning and costs. To do so,
we chose three image classification datasets and one recommender system dataset.
The image classification datasets ranged in task complexity, number
of examples, and classes. We used the following datasets:
\begin{enumerate}
  \item dermnet \cite{shanthi2020automatic}: a dataset of skin images,
  where the task was to determine which one of 23 diseases the image belonged
  to. There were 15,557 training images.

  \item flowers-102 \cite{nilsback2008automated}: a dataset of flower
  images, where the task was to classify images into one of 102 flower classes.
  There were 1,020 training images.

  \item \texttt{cars} \cite{krause20133d}: a dataset of car images, where the
  task was to classify cars into one of 196 categories. There were 8,144
  training images.

  \item \texttt{movielens} \cite{harper2015movielens}: a dataset of users
  ranking movies in IMDB. The training set has 6,040 users, 3,706 movies, and
  900,188 ratings.
\end{enumerate}

We estimated the costs of end-to-end verified training using the \snt protocol by
performing the full training procedure and estimating the cost of constructing
ZK-SNARKs of the training run. We used a variety of MobileNet configurations and
hyperparameters for image classification. We fixed the architecture for the
recommender system but varied the hyperparameters.

\begin{figure*}[t!]
  \centering
  \includegraphics{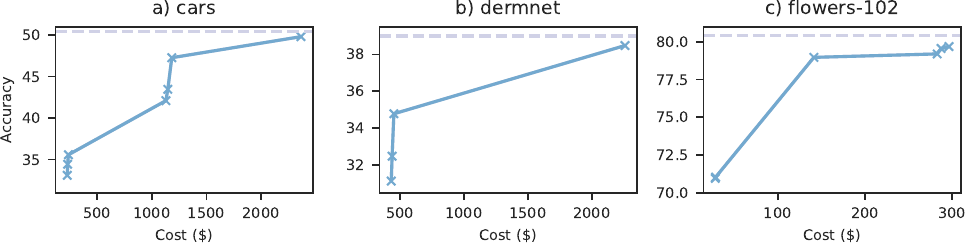}
  \caption{Test accuracy vs cost of proving training across the entire dataset
  for the Pareto frontier of image classification. Higher is better. The dashed
  line is the \texttt{fp32} accuracy.}
  \label{fig:acc-cost}
\end{figure*}

\begin{table*}[t!]
  \centering
  \begin{tabular}{llll}
    Dataset & Accuracy (fixedpoint) & Accuracy (\texttt{fp32}) & Difference \\
    \hline
    dermnet     & 38.5\% & 39.0\% & -0.5\% \\
    flowers-102 & 79.7\% & 80.4\% & -0.7\% \\
    cars        & 49.8\% & 50.4\% & -0.6\%
  \end{tabular}
  \caption{Test accuracy of training with \snt compared to full \texttt{fp32}
  accuracy. The loss in accuracy is marginal across datasets.}
  \label{table:acc-drop}
\end{table*}

\begin{figure}[t!]
  \centering
  \includegraphics[width=0.95\columnwidth]{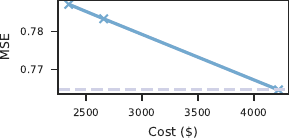}
  \caption{Test MSE vs total training cost for the Pareto frontier for the
  recommender system. Lower is better.}
  \label{fig:acc-cost-movie}
\end{figure}

\begin{figure}[t!]
  \centering
  \includegraphics[width=0.95\columnwidth]{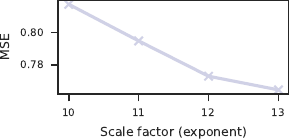}
  \caption{Test MSE vs scale factor. \snt achieves parity with \texttt{fp32} at
  $2^{13}$.}
  \label{fig:acc-sf-movie}
\end{figure}

We show the Pareto optimal frontier of accuracy and cost for the three image
datasets in Figure~\ref{fig:acc-cost} and the mean-squared error (MSE) for the
recommender system in Figure~\ref{fig:acc-cost-movie}. As shown, \snt can
smoothly trade off between accuracy and costs. Furthermore, \snt can achieve high
accuracy on all four datasets despite using fixed-point arithmetic.

Although the costs are high, practitioners can trade off between accuracy and
proving costs. For example, privacy is required in a regulated medical setting, so
the model cannot be revealed. However, for regulatory reasons, the model
provider may desire to provide a transcript of training. In this setting, the
model provider may want to achieve as high accuracy as possible. However, for
some settings where small amounts of accuracy can be traded off for costs, a
model provider can use \snt for as little as \$282 within 1\% of \texttt{fp32}
accuracy. Furthermore, we substantially improve over prior work. Even ignoring
the softmax, the cost of the next cheapest method would be \emph{\$26,637} or
94$\times$ higher.

We further compare the accuracy when using standard \texttt{fp32} training. As
shown in Table~\ref{table:acc-drop}, the accuracy is close to the full precision
counterpart. The recommender system's mean-squared error is on parity with full
\texttt{fp32} accuracy.

\begin{table*}[t!]
  \centering
  \begin{tabular}{llll}
    Model & Accuracy (\texttt{int} division) & Accuracy (rounded, \sn) & Difference \\ 
    \hline
    MobileNet, 0.35, 96  & 59.1\% & \textbf{70.4\%} & 11.3\% \\
    MobileNet, 0.5, 224  & 75.7\% & \textbf{86.3\%} & 10.6\% \\
    MobileNet, 0.75, 192 & 79.2\% & \textbf{88.8\%} &  9.6\%
  \end{tabular}
  \caption{Test accuracy (top-5) of models used by \citet{kang2022scaling} on
  ImageNet with rounded vs integer division. Integer division considerably hurts
  accuracy, indicating worse downstream fine-tuning.}
  \label{table:rounded-div}
\end{table*}

\subsection{Effects of Optimizations}
We investigated the effects of our optimizations: our improved softmax, rounded
division, and precision. To do so, we removed our optimized softmax, removed
rounded division, and reduced the precision (separately) to see the effects.

Removing our optimizations for softmax resulted in failure to train, as the
range of the intermediate values was outside of the feasible range in the
ZK-SNARK. Furthermore, no other work in ZK-SNARKs can perform the softmax,
making training infeasible.

We then changed standard rounded division to integer division (rounded down) and
computed the accuracy of the models on ImageNet as used by
\citet{kang2022scaling}. As shown in Table~\ref{table:rounded-div}, the accuracy
can drop as much as 11.3\%. Since lower accuracy on ImageNet indicates lower
performance for fine-tuning, we used rounded division for all further
experiments.

\begin{figure*}[t!]
  \centering
  \includegraphics[width=\textwidth]{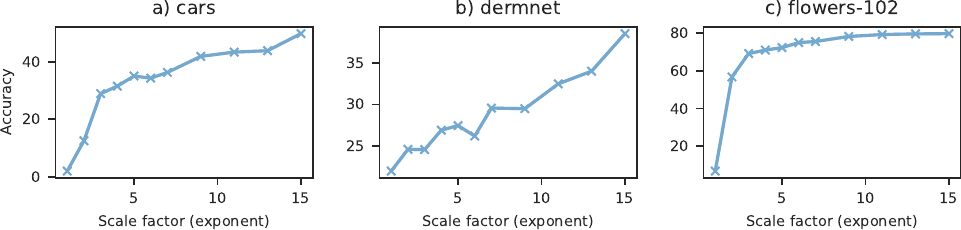}
  \caption{Test accuracy vs scale factor. As shown, we can achieve within 0.7\%
  accuracy compared to full precision with a scale factor of $2^{15}$. The
  accuracy degrades with lower scale factors.}
  \label{fig:acc-sf}
\end{figure*}


We then reduced the training precision from a scale factor of $2^{15}$ to $1$
for the image datasets. We reduced the training precision for the recommender
system dataset from $2^{13}$ to $2^{10}$. We show results for the image datasets
in Figure~\ref{fig:acc-sf} and for the recommender system in
Figure~\ref{fig:acc-sf-movie}.
As shown, the accuracy drops with lower precision. Our results corroborate
results in low-precision training \cite{de2017understanding}.  Nonetheless, \snt
can achieve near-parity with \texttt{fp32} training.

These results show that our optimizations are necessary for high performance in
\snt.

\section{Using \sn for Audits}
\label{sec:audit-eval}

In addition to evaluating the \sn training procedure, we describe and evaluate
end-to-end audits and costs. In principle, \sn is capable of performing any
computable function, but we focus on audits of broader interest. For example, if
the audit simply reveals the training data (which is a computable function), the
model provider may choose to not participate. Our examples are meant as proof of
concepts, and must be combined with work from the transparency literature for a
full end-to-end solution. Our work focuses on the technical feasibility of
privacy-preserving proofs of training and computability of audits. We describe
how to perform an end-to-end audit in Appendix~\ref{apx:twitter}.

Consider the case of recommender systems. Consumers of ML systems are interested
in a wide range of audits. They may be interested in checking if certain items
(e.g., Tweets or products) are censored \cite{pecse2023time}. A more extensive
audit may also attempt to understand \emph{counterfactual} behavior, in which
the inputs to the recommender system are changed, and the effects on the
recommendations are measured \cite{akpinarcounterfactual}. 

Outside of recommender systems, a copyright holder may wish to check that a
model provider did not use their work in training or an auditor may perform a
demographic disparity check. These audits require model outputs and a similarity
check. Each audit requires executing a different function, and as a result has a
different cost profile, which we describe for each audit. We explore these
audits below.

\minihead{Censorship audit}
To perform the censorship audit, we are interested if an item $x$ is ranked
lower than the value implied by the recommender system. In particular, we are
interested in whether an item the user believes should have a high ranking is
censored.

We can use random sampling to determine the quantile of the item $x$ among the
full set of items or a subset of items previously shown to the user (\sni). Determining
the quantile is equivalent to estimating the parameter of a Bernoulli random
variable and has rate $O( 1 / \sqrt{N})$. We use the Hoeffding bound to achieve
finite sample estimates.

We executed this audit on the \texttt{movielens} dataset for estimating the
quantile within 5\% and 1\% (600 and 14,979 samples, respectively). The true
difference in quantile was 1.1\% and 0.1\%, respectively: the Hoeffding bound is
known to be loose in practice \cite{lee2020concentration}. The costs were \$0.42
and \$10.59 for 5\% and 1\%, respectively, which are well within reason for many
circumstances.

\minihead{Counterfactual audit}
The most comprehensive counterfactual audit measures the impact of interventions
on recommender systems \cite{akpinarcounterfactual}. These interventions can be
replacing inputs or changing hyperparameters. In order to do this audit,
we can perform training twice and then estimate quantities. The total cost
is twice the cost of training and the cost of estimating a quantity.

We performed the audit on the \texttt{movielens} dataset. We used a scale factor
of $2^{13}$, which achieves parity with \texttt{fp32} accuracy (see above). The
total cost was \$8,456.

To contextualize this result, the average cost of a financial audit of an S\&P
500 is \emph{\$13,000,000} \cite{audit2022audit}. The full counterfactual audit
would be only \emph{0.07\%} of the cost of a financial audit.


\minihead{Copyright audit, demographic disparity}
In the copyright audit, we prove with ZK-SNARKs that extracted features for each
item (e.g., image) in the training set is dissimilar to features from a
copyright holder's item. For the demographic disparity audit, we computed the
demographic of each item and computed summary statistics. Both audits (from the
perspective of ZK-SNARK computations) are the same cost. We performed these
audit on the \texttt{flowers-102} dataset. The total cost of the audit was \$108
(or about 10 cents per image), showing the feasibility of audits.

\section{Related Work}
\label{sec:rel-work}

\minihead{Secure ML}
Recent work in the ML and cryptography literature has focused on the secure ML
paradigm \cite{ghodsi2017safetynets, mohassel2017secureml, knott2021crypten}.
Much of this work focuses on secure inference, in which a model consumer
offloads computation to a service provider. The model consumer desires either
privacy or validity. The techniques for secure ML range from multi-party
computation (MPC) \cite{knott2021crypten, kumar2020cryptflow, lam2022tabula},
zero-knowledge or interactive proofs \cite{lee2020vcnn, weng2022pvcnn,
feng2021zen, kang2022scaling}, and fully homomorphic encryption (FHE)
\cite{lou2021hemet, juvekar2018gazelle}.

In this work, we focus on \emph{training} as opposed to inference, with
\emph{malicious} adversaries. We provide the first fully-private training scheme
for realistic datasets in the face of malicious adversaries.

\minihead{ZK-SNARKs}
Recent work has optimized ZK-SNARKs for DNNs \cite{lee2020vcnn, weng2022pvcnn,
feng2021zen, kang2022scaling} and numerical optimization problems
\cite{angel2022efficient}. None of the prior work demonstrates how to perform
the softmax and backward pass, both of which are required for training.  In this
work, we leverage ideas from inference to optimize the forward pass but show how
to compute full SGD in ZK-SNARKs and optimize the softmax.

\minihead{Training in ZK-SNARKs}
Recent work has explored training in ZK-SNARKs but have either focused on
smaller models (e.g., logistic regression) \cite{garg2023experimenting} or have
not implemented the softmax (e.g., use the MSE loss to train classification
networks) \cite{sun2023zkdl, abbaszadeh2024zero}. In this work, we implement
algorithms for high-precision softmax in fixed-point and provide algorithms for
end-to-end trustless audits.


\section{Conclusion}

In this work, we present protocols for verifying ML model execution trustlessly
for audits, testing ML model accuracy, and ML-as-a-service inference.
Importantly, these protocols are robust to \emph{malicious} adversaries. We
further present the first ZK-SNARKed ImageNet-scale model to demonstrate the
feasibility of our protocols. Combined, our results show the promise for
verified ML model execution in the face of malicious adversaries. We hope that
future work scales explores scaling to larger models and other domains, such as
language models.



\section{Impact Statement}

This paper presents work whose goal is to advance the field of Machine Learning.
There are many potential societal consequences of our work, none which we feel
must be specifically highlighted here.

\bibliography{paper}
\bibliographystyle{icml2024}

\clearpage
\appendix

\section{ZK-SNARKs}
\label{sec:apx-snarks}

In this section, we describe how general-purpose ZK-SNARK proving systems work.

\subsection{Intuition}
The general intuition behind a ZK-SNARK proving system is that \emph{any}
function can be computed by a polynomial of sufficient size, i.e., that
polynomials are universal. Then, to prove a function, the ZK-SNARK proving
system encodes the function as a polynomial, commits to the polynomial, and
proves that the polynomial evaluation is ``as expected.''

Thus, the primary questions a ZK-SNARK proving system tackles is how the
function is encoded as a polynomial (since different proving restrictions have
constraints on what polynomials can be expressed), how to commit to the
polynomial, and how to prove to the verifier that the polynomial is ``as
expected.''

\subsection{Expressing Functions}
Although ZK-SNARK proving systems fundamentally deal with polynomials (albeit
with specific constraints), functions are expressed via ``front ends.'' These
front ends include R1CS and AIRs. In this work, we focus on AIRs.

As mentioned in Section~\ref{sec:background}, AIRs have a grid of values with
three types of constraints: 1) equality constraints, 2) table constraints, and
3) polynomial constraints. We provide examples of how these can be used to
perform ML operations. Throughout, we denote the number of columns in the grid
as $N$.

\minihead{Sum}
The first example we consider is the sum of a fixed size vector
$\textrm{Sum}(\vec{x}) = \sum_i^n x_i$, where $n = N - 1$. We can lay out the
elements of the vector and the result $z = \textrm{Sum}(\vec{x})$ in a row as
follows:
\[
x_1 | \cdots | x_n | z.
\]
The constraint is:
\[
z - \sum_i^n x_i = 0.
\]

\minihead{Dot product without bias}
Consider a dot product of fixed size without a bias. Namely,
\[
\textrm{DotProd}(\vec{x}, \vec{y}) = \sum_i^n x_i \cdot y_i.
\]
For the gadget, we let $n = \lfloor \frac{N - 1}{2} \rfloor$.

To compute the dot product, we lay out $\vec{x}$ and $\vec{y}$ and the result $z
= \textrm{DotProd}(\vec{x}, \vec{y})$ as follows:
\[
x_1 | \cdots | x_n | y_1 | \cdots | y_n | z
\]
If $N$ is even, we leave a cell empty. The constraint is simply:
\[
z - \sum_i^n x_i \cdot y_i = 0.
\]

Suppose we had two vectors $\vec{x}$ and $\vec{y}$ of cardinality $m > n$. We
can decompose the overall dot product into $\lceil \frac{m}{n} \rceil$ dot
products. We can then use the sum gadget from above to add the partial results.
As we will see, there are many ways to perform a large dot product.

\minihead{Dot product with bias}
Consider a dot product of fixed size with a bias: $\textrm{DotProd}(\vec{x},
\vec{y}, b) = b + \sum_i^n x_i \cdot y_i$. Here, $n = \lfloor \frac{N - 2}{2}
\rfloor$. We can lay out the row as follows:
\[
x_1 | \cdots | x_n | y_1 | \cdots | y_n | b | z
\]
and use the constraint
\[
z - b - \sum_i^n x_i \cdot y_i = 0.
\]

In order to compose a larger dot product, we can decompose the dot product into
$\lceil \frac{m}{n} \rceil$ dot products with biases. The first bias is set to
zero and the remainder of the biases are set to the accumulation. This method of
computing a larger dot product does not require the sum gadget.

As shown in this example, there are a number of ways to perform  the same
operation. The efficiency will depend on a large number of factors, including
the total size of the circuit and the size of the dot products.

\minihead{ReLU}
As a final example, consider computing the ReLU function pointwise over a vector
$\vec{x}$. Here, $|\vec{x}| = \lfloor \frac{N}{2} \rfloor$. We can simply lay
out the row as
\[
x_1 | \textrm{ReLU}(x_1) | \cdots | x_n | \textrm{ReLU}(x_n)
\]

The constraints ensure that pairs of columns $(x_i, \textrm{ReLU}(x_i)) \in T$
for a table $T$ that contains the domain and range of the ReLU function. Other
pointwise non-linearities can be performed similarly.

\subsection{Committing to Polynomials}

Once the function is expressed in the frontend, the ZK-SNARK proving system must
turn the frontend into a series of polynomial commitments. At a high level,
these polynomial commitments (and their openings) consist of the proof. A full
discussion of turning frontend expressions to polynomial commitments is outside
of the scope of this work and we defer to \citet{thaler2022proofs} for an
extended discussion. We discuss the high level ideas here.

There are many forms of polynomial commitments. The two we focus on in this work
are the KZG \cite{kate2010constant} and IPA (inner-product argument)
\cite{bunz2021proofs} commitments.

The KZG commitment scheme requires access to a \emph{structured reference
string} (SRS), which must be generated securely (i.e., with no single party
knowing how the string was generated). This structured reference string is
essentially the powers of a random element in the field, where the random
element is not known to anyone. Fortunately, an SRS has already been generated
for the KZG commitments with 75 active participants \cite{powersoftau}. 

The IPA commitment scheme is transparent, meaning that there need not be a
trusted setup. However, it results in larger proofs. As a result, we focus on
the KZG commitment scheme in this work, but note that IPA works as well
depending on the application at hand.

\section{Protocol}
\label{apx:protocol}

\subsection{Removing Verified Randomness}
\label{apx:no-verified-random}

In order to remove verified randomness, we further require the assumption that
the hash function acts as a random oracle \cite{bellare1993random}. Under this
assumption, the prover first computes a Merkle tree \cite{merkle1988digital} of
the committed data using the random oracle hash function. The root of the Merkle
tree then acts as random bits. To compute more random bits, the prover can
repeatedly hash the Merkle root.

\subsection{Preventing Traversal Ordering Attacks}
Although \sn does not prevent data poisoning attacks, \sn can prevent traversal
ordering attacks \cite{goldwasser2022planting}, in which a model undetectably
changes its behavior on specific inputs. In order to do so, we force the model
trainer to commit to using verified randomness in the future. Under the random
oracle model (see above), we can also use the Merkle root to set the data
traversal ordering.

\subsection{Formal Security Analysis}
\label{apx:formal-analysis}

We provide a formal security analysis of \sn, showing that \sn is knowledge
sound and zero-knowledge. \sn requires two cryptographic primitives: a ZK-SNARK
with standard properties (computational knowledge soundness, zero knowledge) and
a random oracle hash function. We refer to \citet{bitansky2017hunting} for an
extended discussion of ZK-SNARKs and \citet{menezes2018handbook} for an extended
discussion of hash functions.

Denote the verifier to be $V$ and the set of accepting inputs and witnesses to
be $(x, w) \in \mathcal{R}$. The property of computational knowledge
soundness guarantees that for every non-uniform probabilistic polynomial time
(NUPTT) adversary $\mathcal{A}$, there exists an NUPTT extractor
$\textrm{ext}_\mathcal{A}$ such that for all $\lambda$
\[
P\left[ (x, \pi) \leftarrow \mathcal{A}, w \leftarrow \textrm{ext}_\mathcal{A} :
(x, w) \not\in \mathcal{R} \land V(x, \pi) = 1 \right] =
\textrm{negl}(\lambda)
\]
where we omit the dependence on the shared setup for brevity
\cite{atapoor2019simulation}. We use a strong version of zero knowledge called
perfect special honest-verifier zero knowledge (PSHKZK) \cite{halo2}.  Namely,
for two NUPTT adversaries $\mathcal{A}_1, \mathcal{A}_2$, there exists a
polynomial time simulator $\mathcal{S}$ such that
$$P\left[ \mathcal{A}_1(x, \textrm{tr}) = 1 |
  (x, w) \leftarrow \mathcal{A}_2, \mathrm{tr} \leftarrow
  \langle \mathcal{P}(x, w), V(x, \rho) \rangle
\right]$$
$$= P\left[ \mathcal{A}_2(x, \textrm{tr}) = 1 |
  (x, w) \leftarrow \mathcal{A}_2,
  \textrm{tr} \leftarrow \mathcal{S}(x, \rho) \right]
$$
where tr is the transcript and $\rho$ is the internal randomness of the
verifier.

\minihead{Knowledge soundness of \sn}
We first consider if it is possible for the model provider to produce a
transcript of \sn that is invalid. Under the random oracle hash function
assumption, all outputs from the hash function are random. As such, the model
trainer cannot modify the lexicographical ordering of the hashes. Then, the
model trainer produces $T$ ZK-SNARKs, one for each step of the SGD. The
soundness error is amplified by $T$ by the union bound. However, $T \cdot
\textrm{negl}(\lambda)$ is still $\textrm{negl}(\lambda)$. Thus, the entire
protocol is knowledge-sound.

\minihead{Zero-knowledge of \sn}
We then consider if it possible for the verifier to extract information from the
proofs and hashes. Since we use the random oracle assumption, it is
computationally infeasible for an attacker to learn anything about the
pre-images of the hash of the model's weights, which is the only public
information in \snt. To avoid dictionary attacks against the hash, we use a
random salt (i.e., blinding factor) to the model weights.

Furthermore, since we do not modify the underlying protocol of halo2, we inherit
its zero-knowledge property. Namely, a computationally bounded adversary can
learn nothing about the private inputs from the publicly revealed data.

\minihead{Knowledge soundness to bits of security}
When deployed in practice, we must use concrete instantiations of ZK-SNARKs and
hash function. In particular, hash functions do not satisfy the random oracle
model in practice. When performing an analysis of the ``bits of security'' of
ZK-SNARKs and hash functions, we are interested in the computational soundness
of the protocols, since the hashes and proofs are published. 

Let the hash function have $\lambda_1$ bits of security and the ZK-SNARK
protocol have $\lambda_2$ bits of security. Let $\lambda = \min(\lambda_1,
\lambda_2)$. We are interested in the worst case analysis where a single
compromised hash or ZK-SNARK proof compromises the entire protocol. By the
union bound, the bits of security after $D + 4T$ hashes and proofs is $\lambda /
(D + 4T)$.

In practice, $D < T$, so we can upper bound the amplification by $5T$.
Furthermore, $T$ is often in the order of millions, which decreases $\lambda$ by
around 25 bits. As such, to achieve 100 bits of security of $\sn$, we require
$\lambda > 125$. In practice, halo2 has 128 bits of security and the Poseidon
hash function (which we can use) has parameters for 128 and 256 bits of security.

\section{Further Optimizations}
\label{apx:further-opts}

In addition to the optimizations described in the main text, we describe several
additional optimizations to improve performance.

\minihead{Packing}
The maximum size of a lookup table in a halo2 circuit is bound by the number of
rows is the circuit. As a result, if the size of the lookup table is larger than
the size necessary for a single backwards pass, we can pack multiple backward
passes in a single circuit. This is particularly useful for circuits with high
scale factors. We implemented this optimization for all circuits used in the
evaluation.

\minihead{Folding}
Since SGD repeats the same computational pattern many times, we can use
\emph{folding} schemes for proving ZK-SNARKs \cite{kothapalli2022nova}. Folding
schemes are especially useful for repeated computations. Although we have not
implemented this optimization, we expect the costs to be reduced by 10$\times$
or more.

\section{Additional Evaluation Information}

\subsection{Evaluation Setup}
\label{apx:eval-setup}

\minihead{Recommender system model}
We use the following model for the recommender system:
\begin{enumerate}
  \item An embedding dimension of 100 for both movies and users.

  \item The embeddings are then concatenated.

  \item The concatenated embeddings are fed through two fully connected layers.
  The first fully connected layer has an output dimension of 128 and the second
  has an output dimension of 1. The first fully connected layer has a relu6
  nonlinearity.
\end{enumerate}

We used the mean-squared error loss.

\minihead{Hardware}
We use the Amazon Web Services (AWS) \texttt{g4dn.8xlarge} instance type for all
experiments. We use the cost of a spot instance (\$1.286 per hour at the time of
writing) to estimate costs.

\minihead{Code}
We have anonymized our code here: \url{https://anonymous.4open.science/r/zkml-72D8/README.md}

\section{Further Evaluations}

We further evaluated \snt on CIFAR-10 and MNIST. As a comparison, we took the
models from \citet{rathee2023secure}. \citet{rathee2023secure} is in the
\emph{semi-honest} multi-party computation setting, which does not handle
malicious adversaries. Furthermore, it does not produce audits.

We compared the cost producing the proofs for \snt to \emph{just} the bandwidth
costs for \citet{rathee2023secure}. For CIFAR-10 HiNet, \snt costs \$2,417
compared to \$475,234 for \citet{rathee2023secure}. As we can see, \snt is
196$\times$ cheaper than \citet{rathee2023secure}.

\section{End-to-End Twitter Example}
\label{apx:twitter}

As mentioned, complete audits require combining \sn with verified data access
and other forms of verified computation. For example, in the Twitter audit
example, the auditor must also be convinced the input data is valid and that the
preprocessing of the data to the ML model input format is done correctly. In
order to ensure the end-to-end validity, we can combine \sn with other methods
of verified computation and data access:
\begin{enumerate}
  \item To ensure validity of data access, we can use techniques such as vSQL
  \cite{zhang2017vsql}.
  \item To ensure validity of preprocessing, we can use techniques such as
  RISC-Zero or ZK-SNARKs for general purpose computation \cite{arun2023jolt}.
\end{enumerate}


\end{document}